# Empirical Review of Youth-Employment Policies in Nigeria

Oluwasola E. Omoju[1]
Emily E. Ikhide[1]
Iyabo A. Olanrele[2]
Lucy E. Abeng[3]
Marjan Petreski[4]
Francis O. Adebayo[5]
Itohan Odigie[5]
Amina M. Muhammed[6]

## Abstract

Youth unemployment is a major socioeconomic problem in Nigeria, and several youth-employment programs have been initiated and implemented to address the challenge. While detailed analyses of the impacts of some of these programs have been conducted, empirical analysis of implementation challenges and of the influence of limited political inclusivity on distribution of program benefits is rare. Using mixed research methods and primary data collected through focus-group discussion and key-informant interviews, this paper turns to that analysis. We found that, although there are several youth-employment programs in Nigeria, they have not yielded a marked reduction in youth-unemployment rates. The programs are challenged by factors such as lack of framework for proper governance and coordination, inadequate funding, lack of institutional implementation capacity, inadequate oversight of implementation, limited political inclusivity, lack of prioritization of vulnerable and marginalized groups, and focus on stand-alone programs that are not tied to long-term development plans. These issues need to be addressed to ensure that youth-employment programs yield better outcomes and that youth unemployment is significantly reduced.

## Acknowledgements

This work was carried out with financial and scientific support from the Partnership for Economic Policy (PEP www.pep-net.org) working in partnership with the MasterCard Foundation. The authors are grateful to all PEP mentors and resource persons for technical support, guidance, and valuable comments and suggestions on this paper. We also thank all the focus-group discussion participants and key-informant interviewees for giving us invaluable information during the data collection process.

---

[1] National Institute for Legislative and Democratic Studies, Abuja, Nigeria, corresponding author: shollcy@yahoo.co.uk
[2] Nigerian Institute of Social and Economic Research, Ibadan, Nigeria
[3] Baze University, Abuja, Nigeria
[4] University American College, Skopje, North Macedonia
[5] Federal Ministry of Young people and Sport Development, Abuja, Nigeria
[6] YouthHub Africa, Sokoto, Nigeria

# I. Introduction

According to the National Bureau of Statistics, the youth-unemployment rate in Nigeria was 34.9% in 2020, an increase from 29.7% in 2018 (Federal Ministry of Youth and Sports Development, 2021). Overall unemployment is also high and increased from 2.3% in 2000 to 7.5% in 2015, 20.4% in 2017, 27.1% in 2020, and further to 33.3% in 2020. The proportion of Nigerian youth not in education, employment, or training (NEET) rose from 24.8% in 2011 to 28.1% in 2019, suggesting declining opportunities for social mobility and economic potentials for young people in Nigeria.

Rising insecurity and restiveness in recent years have placed the challenge of youth unemployment at the forefront of policy discussions. High youth-unemployment rates are associated with political instability, violence, insecurity, and other social vices (Okafor, 2011). Therefore, the central and sub-national governments have initiated and implemented several youth-employment programs (hereafter, YEP) over the years. The government expended huge human, material, and financial resources into these YEP, yet the youth-unemployment rate has continued to increase rapidly.

In addition, based on data from the Budget Office of the Federation (2023a), funding for the Ministry of Youth and Sports Development has not been consistent. Budget allocation to the Ministry grew by 9.03% in 2019 compared to 2018. The growth rate further increased to 30.22% in 2020 relative to 2019. Thereafter, it declined to 9.83% in 2021 and 6.87% in 2022, before slumping to -0.16% in 2023. Furthermore, budget allocation to the Ministry of Youth and Sports Development and the Ministry of Labor and Employment are low, as they account for only 0.89% and 0.24% of total approved expenditures for 2023, respectively. Funding is also significantly smaller than the 1.75% and 3.70% of total expenditure allocated to the Ministry of Humanitarian Affairs, Disaster Management, and Social Development and for social development and poverty-reduction programs (Budget Office of the Federation, 2023b)

Several studies have been conducted on youth-employment issues and policies in Nigeria, most of which focus on the causes and effects of youth unemployment (Ayinla & Ogunmeru, 2018; Olubusoye, Salisu & Olofin, 2023). Given the growth of YEP in recent years, there has also been empirical focus on how youth perceive YEP. Banfield, Nagarajan, and Olaide (2017) assessed the benefits of government YEP initiated during the national Transformation Agenda of 2011-2015. The study used qualitative data gathered through focus-group discussions and key stakeholder interviews in selected states of northern Nigeria.[7] The states selected were conflict-prone and thus allowed for assessing the possibility of exclusion in YEP. Sixty-one percent of youth agreed that the government was not transparent regarding employment-selection processes, and only 26% believed that government employment interventions reduced youth unemployment. About 80% of the respondents asserted that selection processes were biased in favor of young people with political ties. Furthermore, 64% indicated that government youth-employment programs were biased against women.

Azad and Fashogbon (2018) assessed the performance of public works and skills-for-jobs components of the Youth Employment and Social Support Operation (YESSO) using descriptive analysis. Although YESSO is a national program, the study focused on evaluating the impact of the program in Bauchi state, the only state that has benefitted from the two components of the program in a fragile and conflict-laden northeast Nigeria. For the public work program, only 2% of beneficiaries expressed their dissatisfaction with the type of public work engaged in, while 98% revealed that lack of adequate working tools and supervision impeded their progress in the program. Interestingly, 68% had savings from the program channeled to farm and non-farm enterprises. For the skill-for-jobs component of the program, about 94% of beneficiaries expressed their satisfaction about the adequacy of training days and internship placement. Half of the beneficiaries were unable to save from their stipends, while only 33% started other income-generating activities from their savings. After completing training, 43% of beneficiaries did not start jobs with their acquired skills, which is partly attributable to a lack of competence and necessary tools. The remaining beneficiaries used their skills through personal start-ups, cooperation with other graduates, and working with a trainer.

---

[7] The study did not address a specific youth program during the Transformation Agenda but assessed the benefits derived from the compendium of programs that existed during the period.

Some researchers have analyzed the impact of YEP. McKenzie (2015) evaluated the impact of the Youth Enterprise with Innovation (YouWin) program on new start-ups and existing businesses, using a randomized controlled trial and propensity score matching. McKenzie found that about 55% of jobs generated among new entrants and existing businesses three years after the program were attributable to the YouWin program. The study showed an increase of 14%-22% in innovation practices across multiple dimensions among participating businesses (e.g., quality control, processes, pricing, and internet usage). The study did not account for the gender impact of employment generation, and women constituted only 17.6% of the beneficiaries in the first phase of the program.

Adeyanju and Mburu (2020) examined the impact of the Fadama Graduate Unemployed Youth and Women Support Agropreneur Program (hereafter, FADAMA-GUYS) on youth empowerment in the agricultural value chain using propensity score matching. Program participants had a higher empowerment index than non-participants. Ogunmodede, Ogunsanwo, and Manyong (2020) examined the impact of the N-Power Agro Program on youth employment and income generation in the Oyo, Ogun, and Lagos states of Southwest Nigeria. Participation in the program led to an increase in the monthly income of participants by N 30,191.46 (USD $72), on average, which was almost double the minimum wage at the time of the intervention. However, 80% of the beneficiaries were unable to save from their stipends, limiting their potential to expand or diversify.

Despite the abundance of the literature on youth (un)employment in Nigeria, some gaps remain. Only very few of the YEP, (YouWin, FADAMA-GUYS, and N-Power, for example) have been subjected to impact evaluation. Also, little is known about the implementation challenges faced by these YEP and the extent to which lack of coordination and limited political inclusivity influence the allocation of benefits. The latter are the gaps this paper seeks to fill. The objective of this paper is to identify the challenges facing the implementation of YEP in Nigeria and describe how political economy factors and institutional coordination shape the implementation and ultimately effectiveness of the programs. The paper contributes to the literature on youth unemployment in Nigeria in two major ways. It is the first known study that explores the implementation challenges of YEP in Nigeria. It found that regardless of the number of YEP in Nigeria, challenges of implementation such as inadequate funding, lack of inclusion, limited capacity of implementing agency, lack of M&E framework, adoption of a one-size-fit-all approach, and disconnection from the broader macroeconomic framework undermine their effectiveness and impacts. It is also the first study that comprehensively describes how political economy and institutional coordination factors influence the allocation of benefits in youth-employment programs. The interference in YEP by political-interest groups and lack of proper coordination among YEP implementing agencies pose a challenge to the implementation and ultimately to effectiveness.

The paper is organized as follows. Section II provides a summary of the youth labor market in Nigeria. Section III is the contextual framework that provides an overview of the major YEP in Nigeria. The methodological approach and data collection are the focus of Section IV. The results are presented and discussed in Section IV, while Section VI concludes the paper.

## II. Youth and the Nigerian Labor Market: A Snapshot

Young people (15-34 years) account for 46% of the population of Nigeria. Among youth, young women accounted for 52% vs. 48% for young men. Nigerian young people are excluded from full participation in the labor market, however. Only 37% of youth are in full employment (working at least forty hours per week), down from 71% in 2010. Also, youth unemployment increased from 7% in 2010 to 35% in 2020 (see Figure 1). The labor market is not able to absorb the growing youth population. Poor economic outcomes usually translate into slow job growth or mass layoffs of workers, and young people are disproportionately affected due to limited work experience.

Figure 1: Youth Employment, Unemployment, and Underemployment Rates: 2010-2022 (%)

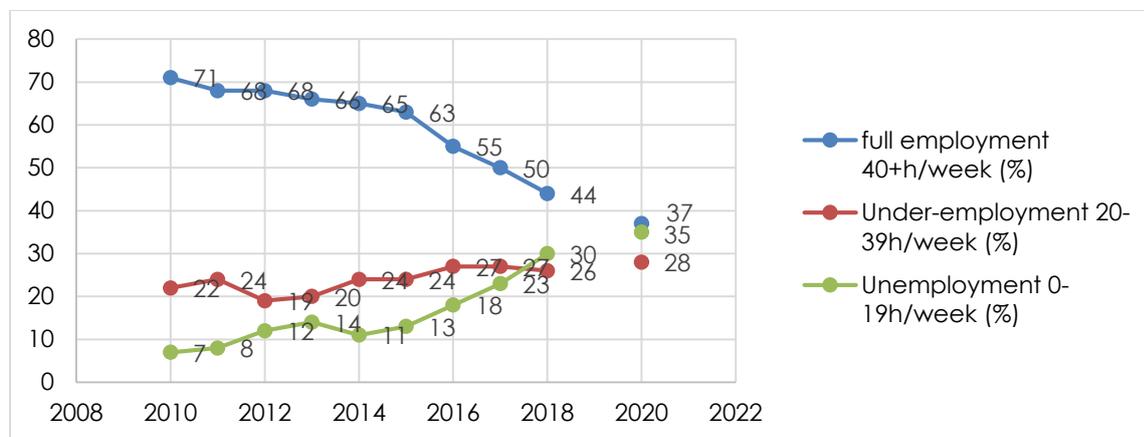

*Source: Federal Ministry of Youth and Sports Development (2021). Note: No data available for 2019; the 2020 values are for Q2 from a computer-assisted telephone interview.*

More young people are unemployed compared to the adult age groups. According to the data from the National Bureau of Statistics, youth (15-34) account for half (forty million) of the total labor force. However, only 65.1% are employed. This is negligible compared to other age groups, as Figure 2 shows. The low participation of youth in the labor market is partly because young people find it difficult to get jobs with their limited experience (Olorunfemi, 2021).

Figure 2: Proportion of Employed and Unemployed by Age Group (%)

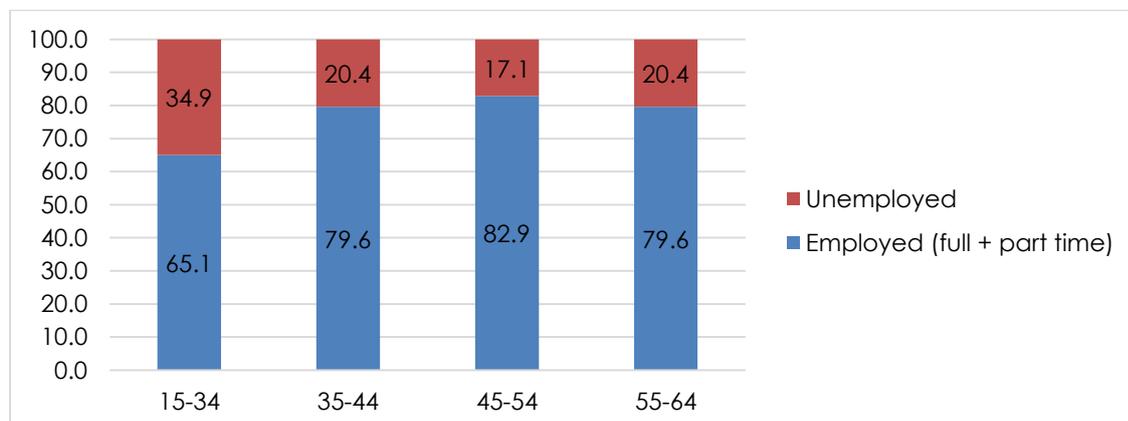

*Source: Computed by authors using data from National Bureau of Statistics (2020)*

Young people also have limited education and training opportunities, thereby undermining their social mobility. In 2011, 24.8% of youth were not in education, employment, or training, as shown in Figure 3. The rates were 19.9% for young men and 30.1% for young women. But the share of youth not in education, employment, or training further increased in 2019, reaching an average of 28.1% (31% of women and 25.3% of men). The larger proportion of women not in education, employment, or training can partly be attributed to the specific challenges women in Nigeria face in access to education and employment. Women largely participate in important but unpaid domestic work, which sometimes hinders them from participating in the labor market (Enfield, 2019).

Figure 3: Share of Youth Not in Education, Employment, or Training (NEET), 2011-2019

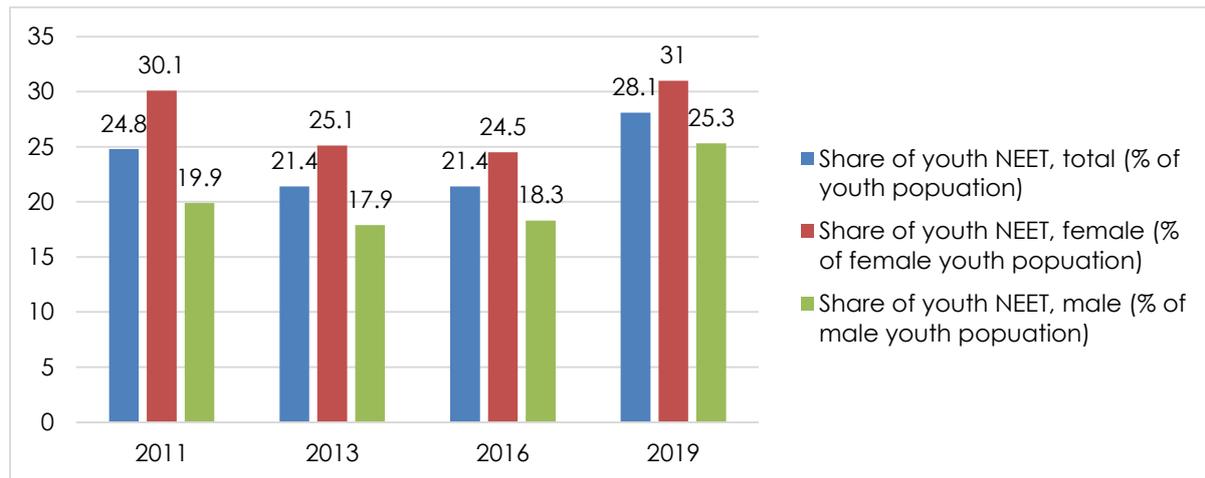

*Source: World Development Indicators, World Bank (n.d.).*

The unemployment problem for youth is often reflected in underemployment in the informal sector, a major sector of the economy. Young people who are unable to get jobs in the formal sector simply enter the informal sector, though the informal sector is characterized by lower wages, absence of structure, lower regulation, and poor working conditions. Underemployment also occurs in the formal sector, particularly if there is a skills mismatch.

Access to quality education is one of the key predictors of labor-market outcomes for youth (Ionescu, 2012). Figure 4 shows that most youth are not in school ("never" and "in school before but not currently enrolled"). Limited access to education among youth also has a noticeable gender dimension. For instance, among the youth group that has have never attended school, 63% are young women compared to 37% who are young men. Similarly, among those who previously attended school but were not currently enrolled, young women accounted for 51%. In contrast, young men accounted for 55% of those in school. This could be explained by the early child marriage culture, particularly in some parts of the country. Early child marriage limits the abilities of young women to attend school and be gainfully employed in the labor market. Finally, only 25% of youth are computer literate.

Figure 4: Youth School Attendance, by Gender, 2020

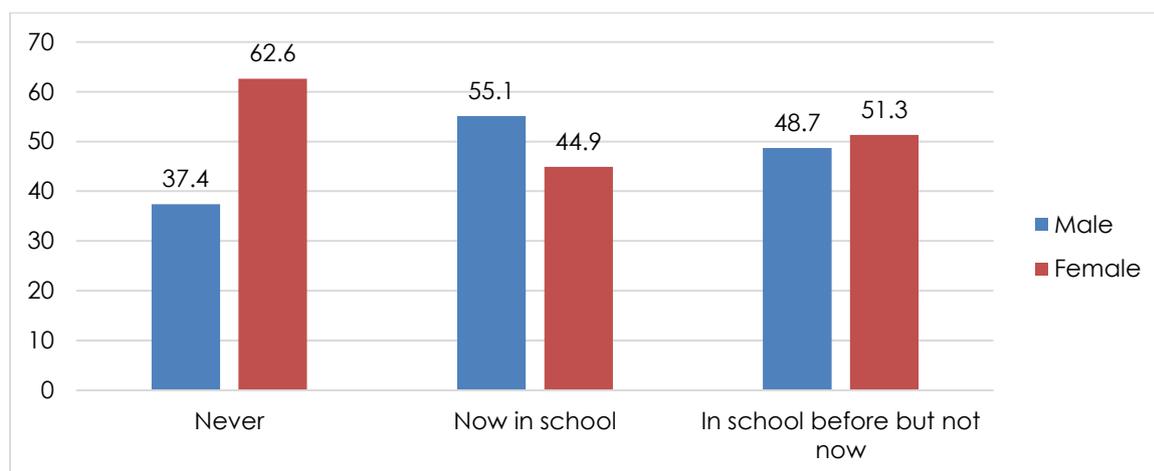

*Source: National Youth Survey 2020 (Federal Ministry of Youth and Sports Development and National Bureau of Statistics, 2020)*

The lack of social and economic mobility for young people in Nigeria has created a grave situation of insecurity and social vices in the country. Several Nigerian young people have also resorted to irregular emigration in order to improve their social and economic livelihoods.

## III. Overview of Youth-Employment Programs in Nigeria

Several youth-employment programs have been implemented in recent years. While the overall objectives of these YEP are to empower young people and enhance their standard of living through dignified employment, the focus, scope, budget, and implementing agency vary from program to program. Some YEP focus on vocational skill acquisition, others are financial support/seed capital, employability skills, and cash-for-job transfer.

Among the YEP that focus on vocational skills are the Skills Development for Youth Employment (SKYE) and the Presidential Youth Empowerment Scheme (P-YES). The SKYE is a needs-based technical and vocational education and training for young people in Nigeria. The objective is to expose young people to different vocational training opportunities through which they can be gainfully employed. The P-YES is also a youth empowerment scheme that focuses on training in vocational skills and government support with the tools needed to function in that vocation over a minimum period of two years. The National Young Farmers Club Program focuses on encouraging youth participation in agriculture. The pilot scheme started with the empowerment of 100 young farmers in animal husbandry (National Agricultural Land Development Authority, n.d.). None of these YEP have been evaluated.

Some YEP also target entrepreneurial skills and financial support for young people. One of these is Youth Enterprise with Innovation (YouWin), a national scheme that provides equity financing to outstanding business plans by small- and medium-scale enterprises of from N 1 to N 10 million (USD $2,380.9-USD $23,809). An evaluation of the program by McKenzie (2015) found that this YEP was responsible for about 55% of jobs created by participants three years after the program, and participants had an increase in innovation practices of about 14%-22% over non-participating businesses. The Youth Entrepreneurship Support Program (YES) also provides business skills (through an extensive eight-week online course and a five-day physical in-class training) and loans of up to N 5 million (USD $11,904) to eligible young people. Other YEP in this category

include the Youth Entrepreneurship Development Program (YEDP) and FADAMA-GUYS. The YEDP is a loan/credit intervention for business proposals for Nigerian young people, offering credit of up to N 3 million (USD $7,142) to eligible youth or N 10 million (USD $23,809) for groups of three to five young people. The Nigerian Youth Investment Fund, on the other hand, makes loans to help young people build and expand their businesses. Beneficiaries receive between N 250,000 (USD $595) and N 3,000,000 (USD $7,142), repayable in five years with an interest of 5% per annum and a twelve-month grace period (Central Bank of Nigeria, 2016). FADAMA-GUYS provides grant support for unemployed young university graduates and women to become agropreneurs. Each participant is given a grant/starter pack of between N 300,000 (USD $714) and N 500,000 (USD $1,190) (Oba, 2019). Adeyanju and Mburu (2020) evaluated the impact of FADAMA-GUYS and found that participants had a higher empowerment index than non-participants.

Other YEP target employability skills through internships or work placement. The N-Power youth empowerment and job creation scheme, launched under the National Social Investment Program in 2016, is one of the most popular. It is a paid, two-year volunteer program (Akujuru & Enyioko, 2019) in which participants receive a monthly stipend of N 30,000 (USD $71) to work in the agriculture, health, or education sectors. An evaluation of the agriculture aspect of the program (N-Power Agro) shows that participants had incomes of about USD $72 more, on average, than non-participants (Ogunmodede, Ogunsanwo & Manyong, 2020). The Jubilee Fellowship Program is a more recent empowerment initiative launched in 2021 by the Federal Government of Nigeria in partnership with the United Nations Development Program. It is an internship program in which young people no more than thirty years old are posted to work in public and private organizations to enable them to acquire skills that enhance their employability.

Cash-for-jobs and direct job programs, which provide temporary employment for young people, also exist. Community Services Women and Youth Employment, a sub-program of the Subsidy Reinvestment and Empowerment Program (hereafter, SURE-P), is one example. This YEP was launched in 2012 to provide temporary employment opportunities for unemployed and unskilled youth and women in labor-intensive communities and sectors (Akande, 2014). The Youth Empowerment and Social Support Operations (YESSO) is also a skill-for-jobs and cash-transfer program targeted at the poor, vulnerable people, and internally displaced people (National Social Safety-Nets Coordinating Office, 2023. The YEP aims to increase the access that poor and vulnerable young people have to employment opportunities. A descriptive evaluation of the YEP by Azad and Fashogbon (2018) showed that 68% of participants were able to save, which was channeled to farm and non-farm activities, and 33% started income-generating activities from their savings. The Special Public Works (SPW) Program is a pilot program by the government aimed at providing direct jobs in rural areas (construction, sanitation, sewage management, etc.). The program was initially designed as a short-term strategy to engage only 1,000 unemployed young people but the coverage has been extended to ease the economic effects of the COVID-19 pandemic (Umar, 2022).

The details of each YEP are provided in Appendix Table 1.

Despite the copious YEP in the country, there has been no deliberate and systematic effort on the part of the government or program implementers to evaluate and ascertain whether programs are effective. The few impact evaluations of YEP that have been done were mainly for academic purposes and were not fed back to the YEP implementer or policy makers for decision making. The lack of rigorous monitoring and evaluation frameworks for YEP is a major challenge that needs to be critically examined.

## IV. Method and Data

Our methodological approaches included desk review, focus-group discussion, and key-informant interviews. The review of YEP was done using desk review. Key-informant interviews and focus-group discussions were also conducted to obtain first-hand information on the employment situation of young people and on youth-employment programs in Nigeria as well as to better understand the political economy of YEP.

We used both primary and secondary data to achieve our objectives. The secondary data were sourced from the literature, surveys, media reports, and government documents. Data collection involved obtaining information about program objectives, commencement and ending dates of program, target groups, beneficiary selection, number of beneficiaries per state, cost of program, and financing. The key-informant interviews were conducted for implementing agencies, key informants, policy influencers, representatives of beneficiary groups, and other relevant stakeholders, and focus-group discussions were conducted with the beneficiaries of youth-employment programs. A purposive sampling technique was used to select institutional participants, including implementing and coordinating ministries, departments, and agencies and other stakeholders relevant to youth-employment programs at the national and sub-national levels.

We employed a multistage sampling technique in selecting youth participants in the focus-group discussions. The first stage involved the selection of one state from each geopolitical zone in Nigeria. Accordingly, the largest state in each of the six geopolitical zones was selected, except for the North East region where we selected Bauchi instead of Borno or Adamawa because of insecurity in the region. Across the states, data collection focused on rural and urban local government areas. Hence, the second stage included the random selection of one rural and one urban local government area. The third stage was the selection of young participants for focus-group discussions, while taking into consideration inclusion of marginalized groups like women, people living with disabilities, and internally displaced persons. Overall, the six selected states include the Federal Capital Territory in the North Central region, Lagos in the South West region, Kano in the North West region, Bauchi in the North East region, Rivers in the South region, and Anambra in the South East region.

A total of eighty key-informant interviews and eighteen focus-group discussions were conducted. Forty of the key-informant interviews were conducted in the Federal Capital Territory because most of the policymakers, influencers and other relevant stakeholders were domiciled there. The remaining key-informant interviews are conducted in Lagos (12), Kano (7), Bauchi (7), Rivers (7), and Anambra (7). A minimum of ten participants were in each focus group. The allocation of the focus-group discussion across the states depended on the population of the state and the strategic nature of the state in Nigeria's political and economic environment. We consequently held five focus-group discussions in Lagos, four in the FCT, three in Rivers, and two each in Kano, Anambra, and Bauchi. Of the 206 participants in focus-group discussions, 44% were women, and 4% were persons with disabilities. Educational levels were: tertiary education, 73%; secondary education, 24%; primary education or less, 3%, and 45% of the participants were employed vs. 55% who were unemployed.

## V. Results

### Challenges to Implementation of YEP in Nigeria

Some of the YEP have accorded young people the opportunity to build entrepreneurial skills that could keep them competitive and relevant in the labor market or assist them in creating livelihood. Some young people noted that the N-power program was able to provide job-relevant skills as well as the ability to create jobs, particularly in the N-Agro, N-Build, N-Creative, and N-Tech sectors. As one focus-group discussion participant in Rivers State said:

> So, I know of a few friends from the youth program that I attended. They actually went on to create their catering businesses and one is a fashion designer and all of these skills were sparked from that skills acquisition training. Not to give the skill acquisition the whole credit but it helped young people to just find somewhere to channel their energy and make something of themselves.

However, the programs are not implemented on a scale sufficient to make a significant impact. YEP targets are low compared to the number of unemployed young people, and implementation usually falls short in reaching actual beneficiaries. This is the view of some of the focus-group discussion participants and youth-civil-society organization (CSOs) leaders. One interviewee stated:

> Also, we look at the youth empowerment programs by the Ministries. We take the numbers of people empowered, and we discovered that, in the region we evaluated, the state with the highest number of (beneficiaries) only had about 5,000 beneficiaries in one year, and that is like a drop in the ocean.

In addition to the actual number of beneficiaries, YEP are also sometimes inadequately inclusive. Although preference is given to women in some YEP (e.g., the SURE-P Community Service Scheme), poor design often leads to the lack of inclusive and equitable opportunities. The YEP rarely have special considerations for diversity across gender or disability. The focus-group discussion participants and representatives of marginalized groups and even some government officials attest to this. A young woman participant in a focus-group discussion in Bauchi State said:

> The selection process is just by luck. There's no special advantage for you being a woman. If you meet the requirement, you are selected, there is no special priority or preference specifically for females.

Meanwhile, the Regional Coordinator for the National Association of Persons with Physical Disabilities, North West Region, noted:

> I know N-Power has benefited some people with disabilities, but I don't know if they were given special consideration. It is not only important for the government to give slots to PWDs, it is mandatory because the disability law (in the country) mandates 2% of all employment opportunities be provided to PWDs, but this hasn't really been practicable. Even the format of sending out the application process needs to be friendly to PWDs.

We also examined whether programs had the ability to implement their mandates. While there was an adequate number of YEP, implementation was usually inadequate. A support officer for a new youth-employment program at one of the UN agencies in Nigeria said:

> I don't think we have a policy problem. Last year, we had the Nigerian Youth Employment Action Plan. That document is detailed, so when it comes to policy I don't think that is the problem. When it comes to programs I am not sure the number of programs is the issue either; we might have to look at how these programs are implemented, and the institutional capacity to implement them.

The YEP are also not tied to a long-term economic development plan, which undermines their sustainability. The lack of connection to a long-term planning framework makes it easy to abandon the programs during government transitions. The changes in area of focus, modes of planning, and budget allocations that come with a new government lead either to a pause or to complete stoppage of existing YEP. In one key-informant interview, an economic expert and advisor to members of the Nigerian Parliament noted that:

> Most of the policies of youth unemployment in Nigeria right now are not derived from a long-term plan; they are ad hoc, and sustainability is an issue. A plan will make the policy/programs sustainable over a long period of time.

Furthermore, YEP are often implemented in isolation or as stand-alone policies. Emphasis is given to youth employment in terms of training, funding, and entrepreneurial skills, but limited attention is given to other aspects of youth development such as behavioral change, mental health, social skills, etc.

Moreover, there is inadequate financial support for program implementation. No clear-cut policy exists regarding resource allocation to YEP because youth-employment policies are not integrated into economic-development and fiscal plans. In addition, the institutional set-up of government agencies in charge of youth development makes funding for YEP opaque. For example, the merger of the ministry in charge of youth and the ministry in charge of sports development undermines transparency in the allocation of funds to the ministry, which mostly go to sports development, rather than youth empowerment. The youth leader of a civil-society organization observed:

> The state government is actually not committed to youth development as they claim. We know this by how much the government is allocating to youth development. For instance, in a particular state, the budget for the Ministry of Youth is about N 600 million (USD $1.4 million), and then only N 200 million (USD $476,190) or even less is going for youth empowerment. (Instead), 70% of the budget is going for sports development.

Moreover, the design and development of YEP do not create effective monitoring and assessment frameworks. Lack of proper definition or absence of performance indicators creates an avenue for weak monitoring and evaluation during and after the YEP. The limited involvement of civil-society organizations in the design and implementation of YEP also stifles accountability and weakens the ability to gauge the success of programs. Furthermore, the limited control that program implementers have over the selection of beneficiaries as a result of lack of political inclusivity often hinders effective monitoring and evaluation. The youth leader of one a civil-society organization put it this way:

> The government doesn't do evaluation of these programs, and that is a concern for development. The whole process is not transparent. For instance, at the end of the whole program, there is no publication that shows the amount of money spent (or) the number of beneficiaries. We don't see all these things released publicly for scrutiny.

The one-size-fits-all approach to YEP across distinct parts of the country is a key issue. Most federal-government-led YEP are designed and implemented with limited consideration of the uniqueness of local economies and demography or of the skills required in each state/region. Cultural orientation, literacy level, belief system, and approach to entrepreneurship differ from state to state.

**Coordination and Political Inclusivity in the Implementation of YEP**

The design and implementation of YEP would ideally involve layers of institutional coordination. The Ministry of Youth and Sports Development is responsible for policy formulation and implementation on issues related to youth development, and the mandate of the Ministry of Labor and Employment is employment-policy formation, review, and implementation in Nigeria. As a consequence of the nature and objectives of most YEP, design and implementation require a coordination mechanism across ministries, departments, and agencies. For example, Youth Enterprise with Innovation program (YouWin) was a collaborative effort of the Ministry of Finance, the Federal Ministry of Youth, the Federal Ministry of Women Affairs and Social Development, and the Federal Ministry of Communication Technology. The Nigerian Youth Investment Fund was designed by the Ministry of Youth and Sports Development, in collaboration with the Nigeria Incentive-Based Risk Sharing System for Agricultural Lending, a financial institution owned by the Central Bank of Nigeria.

A proliferation of YEP developed by various government agencies has nonetheless occurred, though their mandates are not related to youth empowerment, and the result has been the creation of administrative and resource allocation constraints. In some cases, the implementing agency does not engage with the agencies in charge of youth or employment. In addition, coordination between YEP at the federal and state levels is limited, and programs of national coverage have only ad-hoc coordinating bodies at the sub-national levels. The poor coordination among the agencies responsible for YEP is well recognized and aptly captured below in the thoughts of a YEP Program Manager at the Directorate of Youth Empowerment:

> The reason for establishing the "Directorate" is to centralize [all] empowerment programs within the state. Even if the Ministry of Women's Affairs or the Ministry of Youth, or any other agency does an empowerment program, there must be a representative from here so that we will take the data on what they have done, but there is lack of political will for this coordination.

Program coordination is also often reflected in the wide gap between program designers and implementers. Some YEP are designed by experts in government agencies or external consultants but civil servants implement them. Sometimes, designers' ideas regarding the functioning of the programs are not accurately implemented, resulting in wide variations. The deputy director of a YEP implementing agency noted that "most of these programs are developed and launched by this Ministry but were taken away to the Ministry of Finance."

A major issue with the programs is the limited political inclusivity in the implementation, especially in the selection of the beneficiaries. The selection of participants is sometimes influenced by politics. It is common for selection of participants to be done by political party officials or for the committee in charge of the process to be composed largely of political party affiliates. Hence, the programs are often seen as compensation for political party supporters or members, rather than as a YEP. This is confirmed by several key informants and focus-group discussion participants. One from Kano recalled that:

> Like in the case of (a particular YEP), several people from the home state of the minister in charge of the YEP benefitted. The participants provide their information and immediately receive a message that they are successful…. I also benefited from this type of thing. I collected N 20,000 (USD $48) from the local government, but I cannot remember the name of the program. I did not apply. It is a politician from my constituency that submitted my name.

The youth leader of a civil-society organization added that:

> We have (contacted) the government to complain about the selection criteria. The selection criteria are not usually made open. So it is more or less to compensate political partners. The names are mostly gotten from politicians. There is no publicity, there is no advertisement for people to apply for some of these programs. They get their political actors at the local government to arrange certain number of people for empowerment. Then you realize that some programs are targeted at certain people to compensate them for supporting the government.

Meanwhile, poor program coverage is partly due to the concentration of a few participants with political connections. Because of the perceived political influence in the selection process, some young people do not apply for the programs at all. In addition, it is common for participating young people not to use the skills after the YEP because they were not genuinely interested but were selected because of their political connections. A YEP program manager at the Directorate of Youth Empowerment in one of the states explained that:

> According to the plan, young people in every local government are selected through their district heads. But the whole selection process has changed. It is just now based on political interest. The politicians are now the ones selecting the people whom they wish. About 80% of the participants (come through) politicians. Less than 20% are selected on merit.

YEP are often influenced by international donors and organizations. The implementation of these policies designed either by local or international donors is done under the supervision of the Nigerian Federal government. For instance, the Skills Development for Youth Employment (hereafter, SKYE), is sponsored by the German Federal Ministry for Economic Cooperation and Development and the Swiss Agency for Development and Cooperation and executed in partnership with the Nigerian Federal Ministry of Finance, Budget, and National Planning. The Nigerian government finances some of the YEP while the rest are financed by international donors or organizations aligned with their areas of interest. For instance, returnee immigrants are one of the beneficiary groups for SKYE, suggesting that the program may also further Germany's foreign-policy goal of discouraging irregular emigration to Europe by providing employment opportunities to young people who can then earn their livelihood at home.

## VI.  Conclusions and Policy Implications

This paper reviews recent YEP in Nigeria and critically evaluates implementation challenges and the influence of limited political inclusivity on the allocation of YEP benefits in Nigeria. Based on a desk review and analysis of the data and information collected from both primary and secondary sources, it can be inferred that Nigeria's copious youth-employment policies and programs have largely been ineffective in addressing the challenge of youth unemployment. YEP focus on skills acquisition, entrepreneurial training, financial support, and short-term employment schemes. The implementation of these programs usually deviates from the design in terms of funding and in the selection and number of beneficiaries. Some programs do not have strong monitoring or evaluation systems to ascertain the effectiveness of the program. Less than one-fourth of YEP over the years have been subjected to rigorous impact evaluation. Accurate data on budgets and beneficiaries of these programs, moreover, are not publicly available, undermining effective oversight and impact evaluation.

We are able to note several factors that undermine the implementation of YEP in Nigeria, including lack of proper governance and coordination of youth-employment policies, limited scale of YEP, inadequate funding, lack of institutional capacity to effectively implement programs, inadequate oversight of implementation, limited political inclusivity, lack of prioritization of vulnerable and marginalized groups, focus on stand-alone programs that are unconnected to long-term development plans or economic trajectory, lack of collaboration among the relevant stakeholders in the design and implementation of the programs, and weak monitoring and evaluation.

To improve the implementation and effectiveness of YEP and ultimately address the challenges of youth unemployment in Nigeria, we recommend that YEP be aligned to broader long-term national social- and economic-development plans. The governance and coordination of YEP could also be improved to replace the current system in which YEP are implemented by a number of organizations. We recommend that a central coordinating body for YEP be established. Effective oversight systems and monitoring and evaluation frameworks should be established to ensure that the programs are implemented according to design and with limited deviations, and funding must be increased for YEP that yield more benefit relative to cost. Similarly, information about the budget and expenditure of the YEP as well as about beneficiaries should be made publicly available to aid oversight. This can be done by mandating the posting of YEP data and information on the government website and submission of implementation reports to parliament. Youth groups and civil society organizations have a key role to play in this regard. Adequate funding may also be provided for the implementation of YEP. To limit the strain on government finances, YEP implementers may seek the support of the private sector, philanthropists, and international development organizations. Designers of YEP should make efforts to enhance the capacity of the implementers to ensure effective implementation. Lack of political inclusivity in YEP implementation may also be significantly reduced by involving youth groups, civil-society organizations, and other non-governmental stakeholders in the design and implementation. Building institutional capacity can be done through training, collaboration, stakeholder engagement and knowledge sharing. Lastly, YEP should recognize and prioritize vulnerable and marginalized groups, including women, PWDs and IDPs, by imposing quotas to ensure their inclusion.

Appendix 1

Table 1: Summary of YEP As Designed (Targets) and Actual Implementation (Achievement)

| | |
|---|---|
| YEP | The Subsidy Reinvestment and Empowerment Program (SURE-P) |
| Type A=skills/training; B=seed capital; C=job placement/matching; D=subsidies | A, C |
| Key responsible institution | Federal ministries of Finance and Labor & Productivity |
| Brief Definition | Temporal job employment in labor intensive communities and sectors |
| Target | Youth and Women, vulnerable |
| Target Number of Beneficiaries | 185,000 |
| Actual beneficiaries (last year) | 123,049 |
| Selection Criteria | Men Between 18 and 35; Women between 18 and 50; Secondary school graduate; Poor with nothing to do. |
| Geographic Coverage | Nationwide |
| Coverage by Sector | Health, Education, Water, and Sanitation, Environment, Infrastructure Construction and Maintenance, Social Services, and Transport. |
| Coverage of Vulnerable Groups (incl. quotas) | 20% allotted to vulnerable and marginalized group |
| Planned Budget | N/A |
| Amount of Budget Used in Previous Year | N 16.3 billion (USD $41.2 million) |
| Level of Implementation | SURE-P ran from 2012-2014, across all states of the country and achieved most of its objectives. |
| | |
| YEP | Youth Enterprise with Innovation (YouWin) |
| Type A=skills/training; B=seed capital; C=job placement/matching; D=subsidies | B |
| Key responsible institution | Ministry of Finance, Federal Ministry of Youth, the Federal Ministry of Women Affairs and Social Development, and the Federal Ministry of Communication Technology. |
| Brief Definition | A national scheme that provides equity finance to outstanding business plans in small and medium scale enterprises with the aim of establishing new businesses and expanding existing businesses. |
| Target | Local graduate with registered businesses. |
| Target Number of Beneficiaries | 13,500 (Federal Ministry of Finance, 2016) |
| Actual beneficiaries (last year) | 13,500 |
| Selection Criteria | Between 18 and 45; university graduate; all-gender |
|  | inclusive |
| Geographic Coverage | Nationwide |
| Coverage by Sector | Entrepreneurship |
| Coverage of Vulnerable Groups (incl. quotas) | None |
| Planned Budget | N/A |
| Amount of Budget Used in Previous Year | N 9.3 billion (USD $22.1 million) |
| Level of Implementation | The program ran from 2013 through 2015, with the launch of three phases to cover for the lapses that occurred at the end of every year (Federal Ministry of Finance, 2016) |
| YEP | Youth Entrepreneurship Support Program (YES) |
| Type A=skills/training; B=seed capital; C=job placement/matching; D=subsidies | A, B |
| Key responsible institution | Bank of Industry |
| Brief Definition | It provides avenues for young people to build capacity and access funds that can be used to implement their potential innovative ideas |
| Target | Youth Entrepreneurs |
| Target Number of Beneficiaries | 2000 annually (Olagunju, 2016) |
| Actual beneficiaries (last year) | 2,500 (first quarter of the programmer) |
| Selection Criteria | Between 18 and 35; university graduate; National Youth Service Corps (NYSC) member |
| Geographic Coverage | Nationwide |
| Coverage by Sector | Entrepreneurship. |
| Coverage of Vulnerable Groups (incl. quotas) | None |
| Planned Budget | N 10 billion (USD $23.8 million) |
| Amount of Budget Used in Previous Year | N/A |
| Level of Implementation | The programmer was launched in 2016, and 2,500 youths were successfully trained in the first quarter of the implementation (Olagunju, 2016) |
| YEP | N-Power |
| Type A=skills/training; B=seed capital; C=job placement/matching; D=subsidies | A, C |
| Key responsible institution | Federal Ministry of Humanitarian Affairs, Disaster Management, and Social Development |
| Brief Definition | To provide a structure for skill acquisitions and short-term job opportunities for young Nigerians between the ages of 18 and 35 years |
| Target | Graduates and Non-Graduates of Tertiary Institutions |
| Target Number of Beneficiaries | 500,000 Annually (Erezi, 2020) |
| Actual beneficiaries (last year) | 500,000 |
| Selection Criteria | Between 18 and 35 |
| Geographic Coverage | Nationwide |
| Coverage by Sector | Education, Healthcare, and Agriculture |
| Coverage of Vulnerable Groups (incl. quotas) | None |
| Planned Budget | N/A |
| Amount of Budget Used in Previous Year | N 421.5 billion (USD $1 billion) (Ajayi, 2020) |
| Level of Implementation | Kickstarter in June 2016 and has engaged over 500,000 youths over a period of 5 years (Erezi, 2020) |
| | |
| YEP (name) | Youth Entrepreneurship Development Program |
| Type A=skills/training; B=seed capital; C=job placement/matching; D=subsidies | B |
| Key responsible institution | Central Bank of Nigeria |
| Brief Definition | Providing loans to young entrepreneurs for startup and expansion purposes for Small and Medium Enterprises. |
| Target | Youth Entrepreneurs |
| Target Number of Beneficiaries | N/A |
| Actual beneficiaries (last year) | N/A |
| Selection Criteria | Based on the pre-qualification assessment done by the Central Bank of Nigeria's Entrepreneurship Development Centers (EDCs). |
| Geographic Coverage | Nationwide |
| Coverage by Sector | Agricultural value chain (fish farming, poultry, snail farming, etc.), mining and solid minerals, cottage industry, ICT, creative industry (tourism, arts and crafts), and any other activity that may be determined by the CBN from time to time Central Bank of Nigeria, 2016) |
| Coverage of Vulnerable Groups (incl. quotas) | None |
| Planned Budget | N/A |
| Amount of Budget Used in Previous Year | N/A |
| Level of Implementation | Launched on March 15, 2016, but no longer ongoing. |
| | |
| YEP | Skills Development for Youth Employment (SKYE) |
| Type A=skills/training; B=seed capital; C=job placement/matching; D=subsidies | A, C |
| Key responsible institution | German Federal Ministry for Economic Cooperation and Development, Swiss Agency for Development and Cooperation, Nigerian Federal Ministry of Finance, Budget, and National Planning |
| Brief Definition | Needs-based technical and vocational education and training for young people in Nigeria to expose them to training opportunities leading to gainful employment |
| Target | Young people, returnees from abroad |
| Target Number of Beneficiaries | N/A |
| Actual beneficiaries (last year) | 26,149 (GIZ, 2022) |
| Selection Criteria | Between 15 and 35. |
| Geographic Coverage | Nationwide |
| Coverage by Sector | Agriculture, construction, fashion, ICT, and hospitality |
| Coverage of Vulnerable Groups (incl. quotas) | None |
| Planned Budget | € 51.9 million (GIZ, 2022) |
| Amount of Budget Used in Previous Year | N/A |
| Level of Implementation | The programmer is to run from 2018 to 2023. As at 2022, about 26,149 Nigerians with vocational skills in different areas ((GIZ, 2022) |
| | |
| YEP | Nigerian Youth Investment Fund |
| Type A=skills/training; B=seed capital; C=job placement/matching; D=subsidies | B |
| Key responsible institution | Ministry of Youth and Sports Development, Central Bank of Nigeria and the Nigeria Incentive-Based Risk Sharing System for Agricultural Lending |
| Brief Definition | To tackle youth unemployment by providing accessible loan facilities to young people, to enable them to build and expand their businesses which would, in turn, create more job opportunities in critical economic and social sectors. |
| Target | Young people Entrepreneurs |
| Target Number of Beneficiaries | 500,000 (NIRSAL Microfinance Bank, 2020) |
| Actual beneficiaries (last year) | N/A |
| Selection Criteria | Between 18 and 35; registered and unregistered business owners. |
| Geographic Coverage | Nationwide |
| Coverage by Sector | Technology/innovation, agriculture and related value chain, green economy and renewable energy sector, manufacturing, hospitality/tourism, construction, logistics and supply chain, healthcare value chain, creative sector and trading and services |
| Coverage of Vulnerable | None |
| | Groups (incl. quotas) |
| Planned Budget | N 75 billion (USD $178.5 million) (NIRSAL Microfinance Bank, 2020) |
| Amount of Budget Used in Previous Year | N/A |
| Level of Implementation | Launched in October 2020, and currently ongoing |
| YEP | Presidential Youth Empowerment Scheme (P-YES) |
| Type A=skills/training; B=seed capital; C=job placement/matching; D=subsidies | A, B |
| Key responsible institution | Office of the Secretary to the Government of the Federation and the Senior Special Assistant (SSA) to the President on Youth and Student Affairs |
| Brief Definition | Empower youth by creating opportunities and the enabling environments to acquire relevant skills and resources to make them productive, thereby reducing poverty and ensuring an economically empowered youth. |
| Target | Young people |
| Target Number of Beneficiaries | 774,000 (Olaniyi, 2020) |
| Actual beneficiaries (last year) | N/A |
| Selection Criteria | Between 18 and 35; basic English language communication skills |
| Geographic Coverage | Nationwide |
| Coverage by Sector | Technology acquisition, agriculture, catering, fashion, ICT, and mobile money |
| Coverage of Vulnerable Groups (incl. quotas) | None |
| Planned Budget | N/A |
| Amount of Budget Used in Previous Year | N/A |
| Level of Implementation | Launched in 2020. Limited information about the implementation of the programmer. |
| YEP | Special Public Works Program (SPW) |
| Type A=skills/training; B=seed capital; C=job placement/matching; D=subsidies | C |
| Key responsible institution | National Directorate of Employment (NDE), Ministry of Finance, Budget and National Planning, Ministry of Labor and Employment |
| Brief Definition | Providing short term direct jobs in rural areas |
| Target | Young people |
| Target Number of Beneficiaries | 774,000 (Olayinka, 2021) |
| Actual beneficiaries (last | 413,630 |
| year) | |
| Selection Criteria | Between 18 and 35 |
| Geographic Coverage | Nationwide |
| Coverage by Sector | Construction, sanitation, sewage management, etc. |
| Coverage of Vulnerable Groups (incl. quotas) | None |
| Planned Budget | N 52 billion (USD $123.8 million |
| Amount of Budget Used in Previous Year | N 24 billion (USD $57.1 million) |
| Level of Implementation | Launched in 2020, and still ongoing |
| | |
| YEP | Nigeria Jubilee Fellowship Program |
| | |
| Type A=skills/training; B=seed capital; C=job placement/matching; D=subsidies | A |
| Key responsible institution | United Nations Development Program |
| Brief Definition | Connect qualified graduates with job opportunities within their expertise for a year. |
| Target | Unemployed graduates in any discipline |
| Target Number of Beneficiaries | 20,000 (Angbulu, 2021) |
| Actual beneficiaries (last year) | N/A |
| Selection Criteria | Recent unemployed undergraduate; NYSC certificate; below the age of 30 |
| Geographic Coverage | Nationwide |
| Coverage by Sector | All (formal) sectors |
| Coverage of Vulnerable Groups (incl. quotas) | N/A |
| Planned Budget | N/A |
| Amount of Budget Used in Previous Year | N/A |
| Level of Implementation | Launched in August 2021 and still ongoing |
| | |
| YEP | Basic Entrepreneurship and Skills Training Program (BEST) |
| | |
| Type A=skills/training; B=seed capital; C=job placement/matching; D=subsidies | A |
| Key responsible institution | Financing Business for Job and Wealth Creation, the Institute of Entrepreneurship Development, Intuit Financing Inc. and the Nigerian government |
| Brief Definition | Offers a six-month training, skills acquisition and job placement to young people in different vocations. |
| Target | Young people between the ages of 18 and 35 |
| Target Number of Beneficiaries | 774,000 |
| Actual beneficiaries (last | N/A |
| year) | |
| Selection Criteria | All Nigerian citizens and residents aged 18-35. |
| Geographic Coverage | Nationwide |
| Coverage by Sector | All vocational sectors |
| Coverage of Vulnerable Groups (incl. quotas) | N/A |
| Planned Budget | N/A |
| Amount of Budget Used in Previous Year | N/A |
| Level of Implementation | Launched in 2021 |
| YEP | National Young Farmers Club Program |
| Type A=skills/training; B=seed capital; C=job placement/matching; D=subsidies | A |
| Key responsible institution | National Agricultural Land and Development Authority |
| Brief Definition | Encourages youth participation in agriculture. Youth empowerment in agriculture. |
| Target | Young people interested in agriculture |
| Target Number of Beneficiaries | 774,000 |
| Actual beneficiaries (last year) | 100 young farmers benefited in the pilot scheme |
| Selection Criteria | Young people interested in farming |
| Geographic Coverage | Nationwide |
| Coverage by Sector | Agriculture |
| Coverage of Vulnerable Groups (incl. quotas) | N/A |
| Planned Budget | N/A |
| Amount of Budget Used in Previous Year | N/A |
| Level of Implementation | N/A |
| YEP | FADAMA-GUYS |
| Type A=skills/training; B=seed capital; C=job placement/matching; D=subsidies | B |
| Key responsible institution | Ministry of Agriculture and the World Bank |
| Brief Definition | Provide grants of between N 700,000 and N 1,000,000 for youth and women into agriculture business |
| Target | Graduate unemployment youth and women |
| Target Number of Beneficiaries | 6,300 |
| Actual beneficiaries (last year) | N/A |
| Selection Criteria | Unemployed secondary/university graduate; 18-35 years old; at least five years of experience in farming |
| Geographic Coverage | 21 States in the first phase |
| Coverage by Sector | Agriculture |
| Coverage of Vulnerable Groups (incl. quotas) | Unemployed young people and women |
| Planned Budget | N/A |
| Amount of Budget Used in Previous Year | N/A |
| Level of Implementation | N/A |
| YEP | Youth Empowerment and Social Support Operations (YESSO) |
| Type A=skills/training; B=seed capital; C=job placement/matching; D=subsidies | A |
| Key responsible institution | National Social Safety-Net Coordinating Office (NASSCO) |
| Brief Definition | Provides skills for job, coaching and cash transfer for vulnerable members of the society. |
| Target | Unemployed youth, poor, vulnerable people, and internally displaced people |
| Target Number of Beneficiaries | 2,807,656 |
| Actual beneficiaries (last year) | N/A |
| Selection Criteria | N/A |
| Geographic Coverage | Nationwide |
| Coverage by Sector | No specific |
| Coverage of Vulnerable Groups (incl. quotas) | Poor, vulnerable, and internally displaced young people |
| Planned Budget | $100 million |
| Amount of Budget Used in Previous Year | N/A |
| Level of Implementation (to be added later in the project) | N/A |